%% file: main.tex
\documentclass[sigconf, nonacm, screen]{acmart}
\usepackage{tikz}
\DeclareMathAlphabet{\mathcal}{OMS}{cmsy}{m}{n}
\usepackage{amsthm}
\usepackage{textgreek}
\usepackage[english]{babel}

\usepackage{mathtools}

\usepackage{amsmath}
\usepackage{array, xspace, xcolor, algorithm}

\usepackage[noend]{algpseudocode}

\usepackage{adjustbox}
\usepackage{subcaption}
\usepackage{xspace}
\usepackage{enumitem}

\newcommand{\WOC}{WOC\xspace}

\begin{document}

\title{\WOC: Dual-Path Weighted Object Consensus Made Efficient }

\author{Tanisha Fonseca}
\affiliation{%
  \institution{Concordia University}
}
\email{t\_fonse@live.concordia.ca}

\author{Gengrui Zhang}
\affiliation{%
  \institution{Concordia University}
}
\email{gengrui.zhang@concordia.ca}

\begin{abstract}
Modern distributed systems face a critical challenge: existing consensus protocols optimize for either node heterogeneity or workload independence, but not both. For example, Cabinet leverages weighted quorums to handle node heterogeneity but serializes all operations through a global leader, limiting parallelism. EPaxos enables parallel execution for independent operations but treats all nodes uniformly, ignoring performance differences. To tackle this problem, we present \textbf{\WOC}, a dual-path consensus protocol that dynamically routes operations into two paths based on their access patterns. Independent operations execute through a \textbf{fast path} that uses object-specific weighted quorums and completes in one network round-trip. Conflicting or shared objects route through a leader-coordinated \textbf{slow path} employing node-weighted consensus. Our evaluation demonstrates that \WOC achieves up to $4\times$ higher throughput than Cabinet for workloads with $>70\%$ independent objects, while maintaining equivalent performance under high contention.
\end{abstract}
\maketitle
\pagestyle{plain}
\input{sections/Introduction}

\input{sections/background}
\input{sections/WOC_Design}

\input{sections/Protocol_Design_and_Algorithm}
\input{sections/Evaluation}
\input{sections/Conclusion}

\bibliographystyle{ACM-Reference-Format}
\bibliography{ref.bib}
\end{document}

%% file: sections/Introduction.tex
\section{Introduction}
\label{sec:introduction}

Distributed consensus protocols form the backbone of modern cloud infrastructure, enabling coordination across replicated databases~\cite{corbett2013spanner,taft2020cockroachdb,huang2020tidb}, blockchains, and distributed storage systems. Traditional algorithms such as Paxos~\cite{lamport1998part,lamport2001paxos} and Raft~\cite{ongaro2014search} rely on uniform majority quorums, treating all replicas equally. While they guarantee strong consistency, they face two limitations in heterogeneous deployments: they fail to leverage differences in replica capabilities, and they serialize all operations through a single leader, preventing the parallel execution of independent updates.

Recent protocols address these limitations individually. Cabinet~\cite{zhang2025cabinet} introduces node-weighted consensus to prioritize faster and more reliable replicas but requires centralized coordination. EPaxos~\cite{moraru2013there} enables parallel execution of independent operations but assumes all replicas are equal, ignoring node heterogeneity.

This dichotomy reveals a significant gap: existing protocols optimize along a single dimension, either node heterogeneity or object independence, but not both. \WOC bridges this gap with a dual-path approach. Independent operations execute through a fast path using object-specific weighted quorums, while conflicting or shared objects follow a node-weighted consensus path. \WOC adapts dynamically to both workload and replica performance, achieving high throughput for low-contention workloads while maintaining correctness under conflicts.

To summarize, \WOC makes the following contributions:

\begin{enumerate}
\item We propose a dual-path protocol that integrates leaderless object-weighted  fast path consensus with leader-coordinated node-weighted consensus (slow path),  dynamically routing operations based on contention and replica availability.

\item We introduce \textit{object-weighted consensus}, extending weighted quorum systems from node granularity to object granularity, enabling fine-grained parallelism for independent operations.

\item We evaluated \WOC across multiple dimensions: batch size, conflict rate, client concurrency, and server scalability, demonstrating consistent performance advantages in low to medium contention scenarios.
\end{enumerate}

The rest of the paper is organized as follows: \S\ref{sec:background} presents the background and motivation; \S\ref{sec:design} describes the \WOC design and object-weighted quorum foundation; \S\ref{sec:protocol} presents the protocol algorithms; and \S\ref{sec:evaluation} reports the evaluation results.

%% file: sections/background.tex
\section{Background and Motivation}
\label{sec:background}

\subsection{Consensus and Fault Tolerance}

Consensus algorithms are a cornerstone of state machine replication (SMR) in distributed systems, enabling a set of replicas to agree on a totally ordered sequence of state transitions despite failures~\cite{schneider1990implementing}. Consensus protocols are typically designed under one of two fault models: \emph{crash fault tolerance} (CFT), where nodes may fail by stopping, and \emph{Byzantine fault tolerance} (BFT), where nodes may behave arbitrarily or maliciously~\cite{zhang2024reaching}. The BFT model assumes a stronger adversary and is adopted by protocols such as HotStuff~\cite{yin2019hotstuff}, Narwhal~\cite{danezis2022narwhal}, Prosecutor~\cite{zhang2021prosecutor}, and PrestigeBFT~\cite{zhang2024prestigebft}. However, due to its substantially lower communication and computational overhead, CFT remains the dominant choice in performance-critical and production systems. 

Classical CFT consensus protocols, such as Paxos~\cite{lamport1998part}, Raft~\cite{ongaro2014search}, and their variants~\cite{marandi2010ring, lamport2004cheap, lamport2009vertical, zhang2018efficient, shi2016cheap}, adopt \textbf{majority quorums}, requiring agreement from at least $\lceil n/2 \rceil + 1$ nodes. This simple majority rule guarantees quorum intersection and has become the de facto standard in practical systems~\cite{agrawal1992generalized}. However, in heterogeneous environments where replicas exhibit substantial differences in processing speed, network latency, or resource capacity, uniform voting can force fast replicas to wait for slower ones, leading to increased commit latency and degraded overall system performance.

Weighted voting, originally introduced by Gifford~\cite{gifford1979weighted}, demonstrated that assigning different vote weights to replicas could optimize read and write quorums. Building on this idea, subsequent work formalized weighted quorum systems and generalized quorum constructions, providing rigorous guarantees on quorum intersection and fault tolerance~\cite{herlihy1986quorum, zhang2022escape}. More recent research revisited quorum design to relax strict majority requirements, including flexible quorum~\cite{howard2017flexible} and vertical Paxos, which enables reconfiguration and vertical scaling by decoupling configuration changes from the normal consensus path~\cite{lamport2009vertical}.

Recently, Cabinet \cite{zhang2025cabinet} extended the concept of weighted voting by introducing \textbf{dynamically weighted consensus}. In Cabinet, each node $i$ is assigned a weight $w_i$ reflecting its observed responsiveness, and the top $t+1$ weighted replicas form a cabinet that drives consensus progress. Instead of requiring agreement from a numerical majority of nodes, Cabinet employs weighted quorums, where a decision is committed once the accumulated weight exceeds a consensus threshold: $CT = \sum w_i / 2$. Two invariants ensure correctness: (a) the sum of the $t+1$ highest weights exceeds $CT$ (progress), and (b) the sum of the top $t$ weights remains strictly below $CT$ (safety).

While Cabinet demonstrates significant performance improvements in heterogeneous deployments, \textbf{Cabinet is workload agnostic and remains fundamentally node-centric}: all operations are funneled through a single global consensus instance coordinated by one elected leader. As a result, even independent or non-conflicting operations are serialized, preventing the system from exploiting inherent parallelism.

\subsection{Object-Aware Consensus Protocols}

EPaxos~\cite{moraru2013there} pioneered a different optimization by observing that operations on independent objects don't need global ordering. Instead of routing everything through a leader, any replica can coordinate operations. Each proposal includes a dependency set identifying which other commands it conflicts with. Commands with non-overlapping dependencies can commit simultaneously, while conflicting commands follow their dependency ordering.

This design works particularly well across geographic regions where clients connect to nearby replicas rather than always contacting a global leader. EPaxos distributes load evenly and achieves optimal commit latency. However, it makes a critical assumption: all replicas are equally capable. Every replica contributes one vote regardless of its actual performance characteristics. In heterogeneous deployments, this egalitarian approach leads to suboptimal quorum formation. A slow or unreliable replica carries the same voting power as a fast, reliable one, potentially increasing commit latency and reducing availability. Moreover, EPaxos applies uniform treatment to all objects: identical quorum sizes and protocol rules regardless of access patterns or contention.

\subsection{Bridging Node and Object Heterogeneity}

We need a solution that accounts for both types of heterogeneity. Therefore, we propose a dual-path system that combines both: a fast path for independent objects using object-specific weights, and a slow path for conflicting or shared objects using a node-weighted consensus.

Consider a multi-tenant database where tenants are geographically distributed. Tenant A's data gets accessed primarily from US replicas, while Tenant B's requests come from European replicas. In \WOC, Tenant A's objects assign higher weights to US replicas, enabling faster quorum formation for Tenant A's operations. Shared system resources use \WOC's slow path with global node weights for proper coordination across regions. This adaptive approach utilizes both node heterogeneity (fast replicas carry more weight) and object independence (tenant-specific data executes in parallel).

%% file: sections/WOC_Design.tex
\section{Weighted Object Consensus Design}
\label{sec:design}

\subsection{Object-Weighted Quorum Foundation}
\WOC applies Cabinet's~\cite{zhang2025cabinet} weighted quorum intersection rules at object-level granularity. For each object $O$ in the system, we define:

\begin{itemize}
\item \textbf{Object weight vector}: $W^O = [w_1^O, w_2^O, \ldots, w_n^O]$ where $w_i^O$ represents replica $i$'s weight for object $O$
\item \textbf{Object consensus threshold}: $T^O = \sum_{i=1}^{n} w_i^O / 2$
\item \textbf{Object quorum}: Any subset $Q^O \subseteq \{1, \ldots, n\}$ satisfying $\sum_{i \in Q^O} w_i^O \geq T^O$
\end{itemize}

\textbf{Dynamic weight assignment.} Object weights reflect access patterns and replica efficiency.WOC dynamically assigns object weights based on replica responsiveness. Specifically, replicas that respond faster to requests for object $O$ receive higher weights for that object. If replica $i$ frequently serves read and write requests for object $O$ with low latency, it receives higher weight $w_i^O$. Conversely, replicas that exhibit slower response times or higher latencies for object $O$ receive lower weights. This weight assignment is updated continuously based on observed response times, ensuring that the fastest-responding replicas for each object maintain the highest weights.

For example, consider an object $O$ where replica $R_1$ consistently responds within 5ms, replica $R_2$ responds within 10ms, and replica $R_3$ responds within 20ms. \WOC would assign the highest weight to $R_1$, a moderate weight to $R_2$, and the lowest weight to $R_3$ for this specific object. Importantly, these weights are object-specific: replica $R_3$ might have high weight for a different object $O'$ if it responds faster for that particular object.

This design enables smaller, faster quorums for objects with localized access patterns while maintaining correctness through threshold-based validation. By prioritizing fast-responding replicas, \WOC minimizes the time required to achieve quorum, as operations can commit as soon as the fastest $t+1$ replicas (those with the highest weights) respond.

For the slow path, \WOC employs global node weights $W^N = [w_1^N, w_2^N, \ldots, w_n^N]$ and node consensus threshold $T^N = \sum_{i=1}^{n} w_i^N / 2$. Node weights reflect overall replica reliability and responsiveness across all objects, enabling efficient coordination for shared resources. Unlike object-specific weights, node weights represent a replica's general performance characteristics across the entire system.

\subsection{Geometric Weight Assignment}
\WOC employs geometric weight distribution with tunable steepness. For an object $O$ with $n$ replicas ordered by decreasing efficiency, weights are assigned as:

\begin{equation}
w_i^O = R^{n-1-i} \text{ for } i \in \{0, 1, \ldots, n-1\}
\end{equation}

where $R \in [1.0, 2.0]$ controls distribution steepness. Low values ($R \approx 1.3$) produce uniform distributions requiring more replicas per quorum, improving fault tolerance. High values ($R \approx 2.0$) create steep gradients where top replicas dominate, enabling smaller quorums but increasing sensitivity to failures.

\begin{table}[ht]
\centering
\caption{Object-weighted geometric distributions with different failure thresholds. Cabinet members (top $t+1$ weighted replicas) are highlighted. $T^O$ represents the object consensus threshold.}
\label{tab:object-weights}
\Description{Table demonstrating how geometric weight distribution enables smaller, faster quorums by prioritizing responsive replicas.}

\begin{adjustbox}{width=0.95\linewidth}
\begin{tabular}{c|c|ccccccc|c}
\textbf{Object} & \textbf{$R$} & \textbf{R1} & \textbf{R2} & \textbf{R3} & \textbf{R4} & \textbf{R5} & \textbf{R6} & \textbf{R7} &\textbf{$T^O$} \\
\hline
ObjA ($t=1$) & 1.40 & \textbf{7.53} & \textbf{5.38} & 3.84 & 2.74 & 1.96 & 1.40 & 1.00 & 11.93 \\
ObjB ($t=1$) & 1.38 & \textbf{6.91} & \textbf{5.00} & 3.63 & 2.63 & 1.90 & 1.38 & 1.00 & 11.23 \\
ObjC ($t=2$) & 1.25 & \textbf{3.81} & \textbf{3.05} & \textbf{2.44} & 1.95 & 1.56 & 1.25 & 1.00 & 7.54 \\
ObjD ($t=3$) & 1.10 & \textbf{1.77} & \textbf{1.61} & \textbf{1.46} & \textbf{1.33} & 1.21 & 1.10 & 1.00 & 4.74 \\
\end{tabular}
\end{adjustbox}
\end{table}
\textbf{Example: } Consider a system with $n=7$ replicas where the fastest replica for object $O$ is $R_1$, followed by $R_2$, $R_3$, etc. Using $R=1.40$ (as in ObjA from Table~\ref{tab:object-weights}):
\begin{itemize}
\item $w_1^O = 1.40^{7-1-0} = 1.40^5 = 7.53$ (fastest replica gets highest weight)
\item $w_2^O = 1.40^{7-1-1} = 1.40^4 = 5.38$ (second fastest)
\item $w_3^O = 1.40^{7-1-2} = 1.40^3 = 3.84$ (third fastest)
\item ... continuing down to $w_6^O = 1.4^0 = 1.0$ (slowest replica)
\end{itemize}

The total weight is $\sum w_i^O = 22.86$, so the consensus threshold is $T^O = 22.86/2 = 11.43$. Since $w_1^O + w_2^O = 12.91 > 11.43$, consensus can be achieved with just the two fastest replicas. This geometric progression ensures that fast replicas can form quorum without waiting for slower one.

\begin{table}[ht]
\centering
\caption{Node-weighted geometric distribution for slow path. Cabinet members (highest weighted nodes) are highlighted for each failure threshold $t$.}
\label{tab:node-weights}
\Description{Table showing global node weight configurations used for leader-coordinated slow-path consensus.}
\begin{adjustbox}{width=0.85\linewidth}
\begin{tabular}{c|c|ccccccc}
\textbf{$t$} & \textbf{$r$} & \textbf{$w_1$} & \textbf{$w_2$} & \textbf{$w_3$} & \textbf{$w_4$} & \textbf{$w_5$} & \textbf{$w_6$} & \textbf{$w_7$} \\
\hline
1 & 1.40 & \textbf{7.5} & \textbf{5.4} & 3.8 & 2.7 & 2.0 & 1.4 & 1.0 \\
2 & 1.38 & \textbf{6.9} & \textbf{5.0} & \textbf{3.6} & 2.6 & 1.9 & 1.4 & 1.0 \\
3 & 1.19 & \textbf{2.8} & \textbf{2.4} & \textbf{2.0} & \textbf{1.7} & 1.4 & 1.2 & 1.0 \\
4 & 1.08 & \textbf{1.6} & \textbf{1.5} & \textbf{1.4} & \textbf{1.3} & \textbf{1.2} & 1.1 & 1.0 \\
\end{tabular}
\end{adjustbox}
\end{table}
\textbf{(Node-Weighted Quorums)} Table~\ref{tab:node-weights} shows global weight distributions used for the slow path when operations conflict across objects. Unlike object-specific weights, these node weights are uniform across all objects:
Each row represents a different failure threshold configuration across a 7-node system. The Cabinet members, highlighted in the table, are the replicas whose combined weight is sufficient to form a quorum. For $t=1$, replicas 1 and 2, with weights 7.5 and 5.4 respectively, together exceed half the total weight, forming the cabinet. As $t$ increases, more replicas join the cabinet, and the weight distribution becomes more uniform. For $t=4$, five replicas with weights 1.6, 1.5, 1.4, 1.3, and 1.2 form a nearly equal-weight cabinet.

Together, the tables demonstrate how \WOC adapts weight distributions to both object-granularity (Table~\ref{tab:object-weights}) and overall node reliability (Table~\ref{tab:node-weights}), balancing fast-path efficiency with slow-path fault tolerance. In practice, \WOC continuously updates these weights based on observed response times, ensuring that the most responsive replicas maintain the highest weights for each object.

\subsection{Object Classification and Adaptive Routing}
\WOC dynamically classifies objects into three categories based on observed access patterns:

\textbf{Independent Objects.} Objects whose operations are isolated to a single client and do not depend on other objects. Examples include a user's personal bank account or an individual's shopping cart. In a distributed system, operations on independent objects can be executed concurrently across replicas using the fast path, leveraging object-specific weighted quorums.

\textbf{Common Objects}
Objects shared among multiple clients, such as a joint bank account, which may experience occasional concurrent access. In distributed deployments, operations on common objects require coordination to preserve consistency, so they are routed through the slow path, employing node-weighted consensus to ensure correctness.

\textbf{Hot Objects}: Objects that experience frequent simultaneous access from multiple clients. Due to high contention, distributed systems must serialize operations on hot objects using the slow path, guaranteeing linearizability across all replicas.

Object Classification adapts continuously. \WOC maintains per-object statistics tracking operation frequency, conflict rates, and access latency, enabling dynamic routing decisions that optimize for both performance and correctness.

%% file: sections/Protocol_Design_and_Algorithm.tex
\section{Protocol Design and Algorithms}
\label{sec:protocol}

\subsection{System Model}
\WOC operates in a partially synchronous distributed system with $n$ replicas and a configurable fault threshold such that $t$ ($1 \leq t \leq \lfloor (n-1)/2 \rfloor$). \WOC assigns dynamic weights to replicas and designates the $t+1$ highest-weight replicas as quorum members, tolerating failures of any $t$ replicas while guaranteeing progress when these $t+1$ replicas remain available. The system manages a set of objects $\{O_1, O_2, \ldots, O_m\}$, each supporting read and write operations. Clients contact any replica to submit requests. Replicas communicate via asynchronous remote procedure calls (RPCs) with eventual delivery guarantees.

\subsection{System Architecture}
\begin{figure}[t]
    \centering
    \includegraphics[width=0.99\linewidth, height=0.30\textwidth]{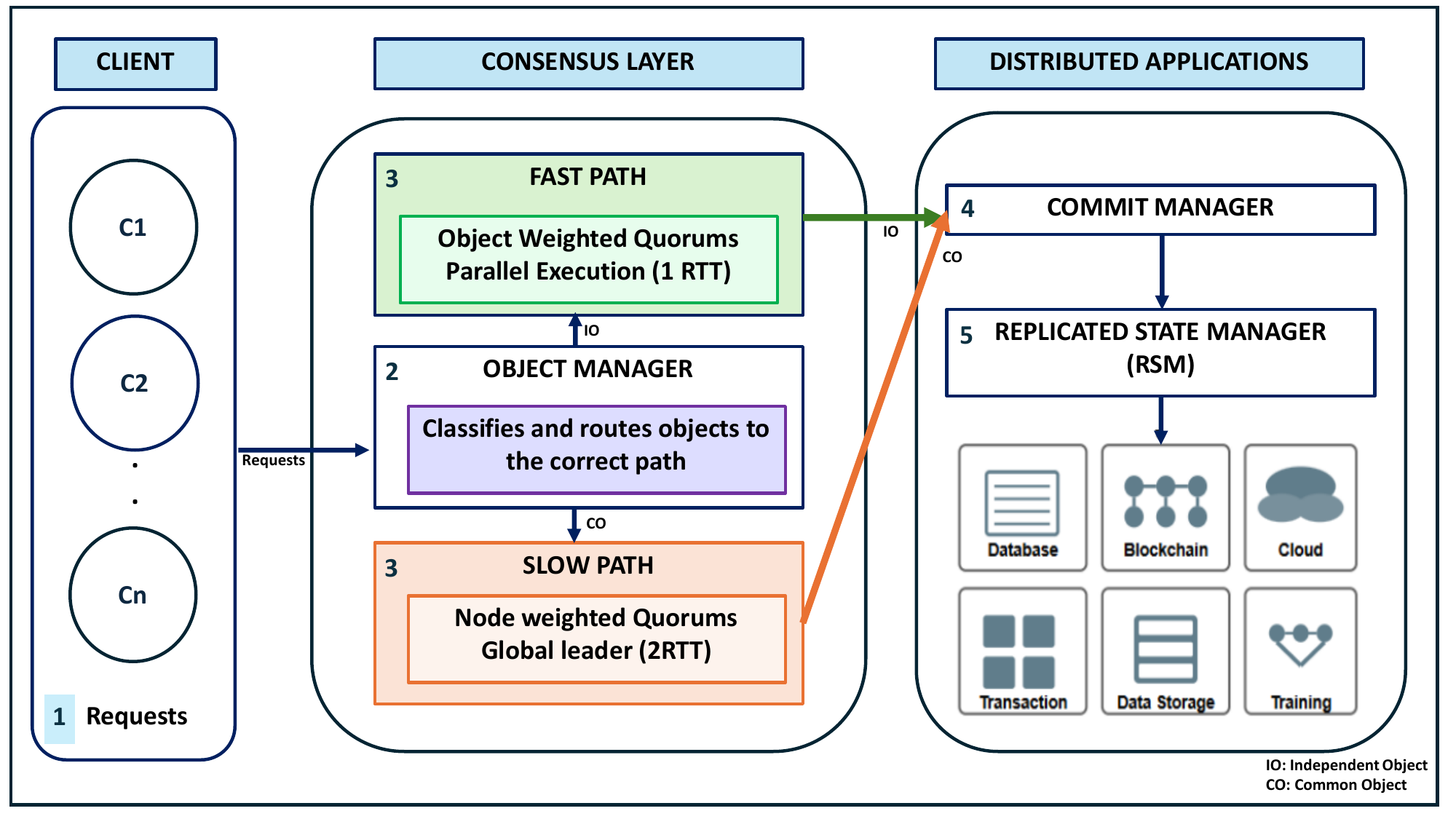}
    \caption{WOC system architecture showing three-layer design: client layer for request submission, consensus layer with Object Manager routing requests to either fast path or slow path, and distributed applications layer with RSM for consistent state replication.}
    \label{fig:architecture}
    \Description{Architecture diagram showing WOC's three layers with arrows indicating request flow from clients through the consensus layer's dual paths to the replicated state machine.}
\end{figure}

Figure~\ref{fig:architecture} illustrates the \WOC architecture, which consists of three main layers: the client layer, the consensus layer, and the distributed applications layer. The consensus layer is the core of \WOC's design, implementing the dual-path routing mechanism that enables efficient handling of heterogeneous workloads.

\textbf{Client Layer.} Multiple clients submit requests to any available replica in the system. Unlike traditional leader-based protocols where clients must discover and contact a specific leader, \WOC allows clients to connect to their nearest or most responsive replica. This design reduces client-perceived latency and distributes the ingress load across the cluster. Each client sends operation requests specifying the target object and the desired state transition.

\textbf{Consensus Layer.} The consensus layer consists of the Object Manager, Fast Path, and Slow Path. The Object Manager maintains metadata about each object, classifying them as Independent (IO), Common, or Conflicting (CO), tracks in-flight operations, and routes requests accordingly. Independent objects with no conflicts go through the fast path, which uses object-specific weights to form quorums and achieve consensus in a single round-trip, enabling parallel, leaderless execution and high throughput. Common and conflicting objects are routed to the slow path, where a leader coordinates operations using global replica weights to ensure correctness and linearizability, potentially requiring two round-trips but allowing dynamic reordering of non-conflicting operations within the same batch to improve efficiency.

\textbf{Distributed Applications Layer.} Both fast and slow paths commit operations via the Commit Manager to the Replicated State Machine (RSM), which applies them consistently across replicas. \WOC supports a variety of distributed applications, including transactional databases, ledgers, file systems, key-value stores, caches, and machine learning parameter servers. By separating concerns—Object Manager for routing, fast path for independent operations, and slow path for conflicting operations—\WOC dynamically adapts to workload contention, achieving high throughput for low-contention workloads while ensuring correctness under high contention.

\subsection{Fast Path: Object-Weighted Consensus}

The fast path is designed for \textbf{independent objects} and operates without global leader coordination, enabling parallel execution. Algorithm~\ref{algo:fastpath} outlines the complete fast-path protocol. Operations follow these steps:

\begin{enumerate}
\item \textbf{Conflict Detection:} A coordinator checks if the object $O$ has in-flight conflicting operations (lines 2-3). If so, the operation is routed to the slow path.

\item \textbf{Proposal Broadcast:} The coordinator sends a \texttt{FAST\_PROPOSE} message to all replicas (lines 5-7).

\item \textbf{Weighted Voting:} Each replica responds with \texttt{FAST\_ACCEPT} if no conflicts exist, or \texttt{CONFLICT} otherwise (lines 10-11).

\item \textbf{Early Termination:} The coordinator accumulates weights from accepting replicas (line 11). Once the object quorum threshold $T^O$ is reached, the operation commits immediately (lines 12-13).

\item \textbf{Commit Broadcast:} \texttt{FAST\_COMMIT} is sent to all replicas (line 13), which apply the operation locally.
\end{enumerate}

\begin{figure}[b]
    \centering
    \includegraphics[width=0.99\linewidth]{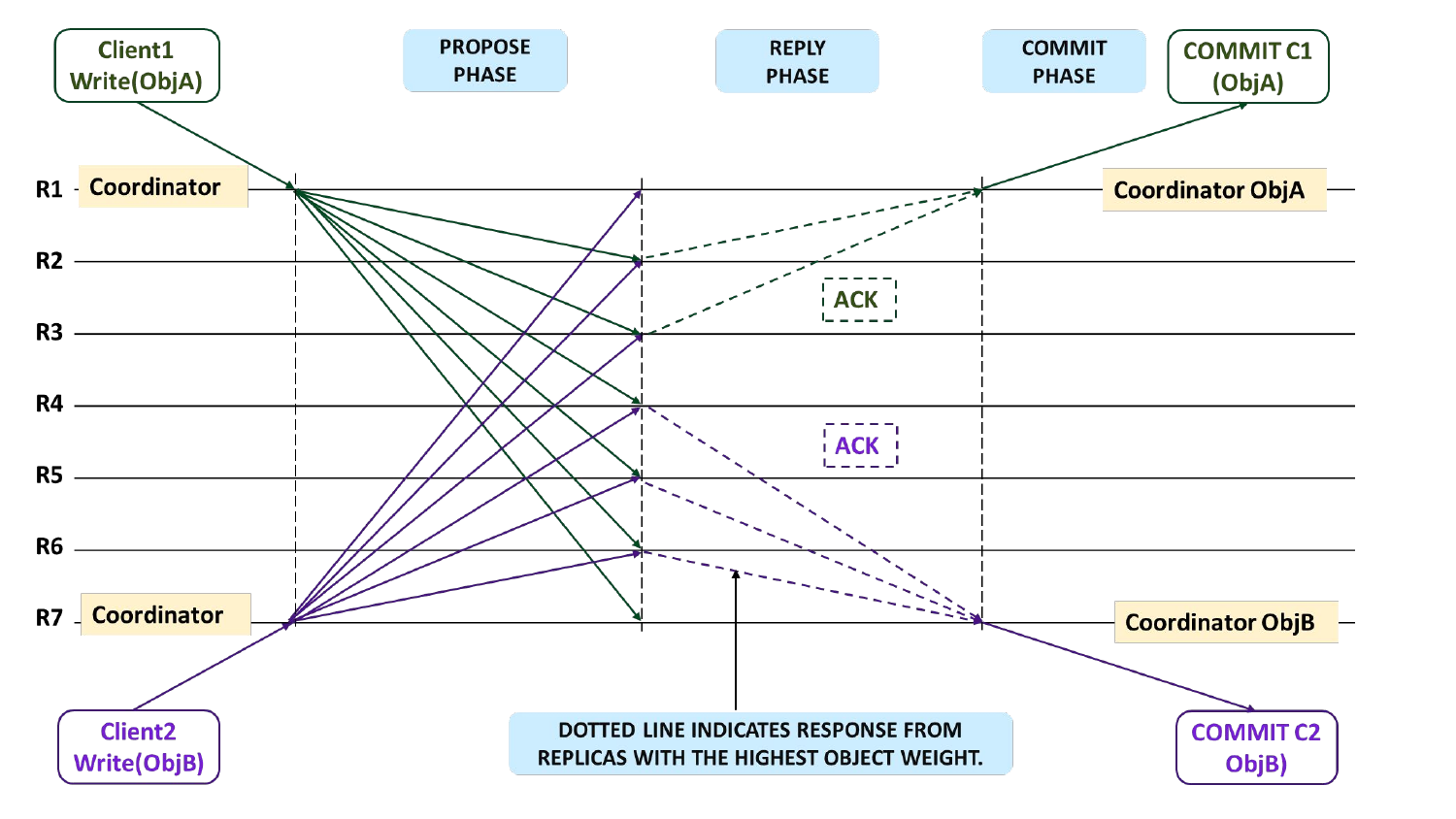}
    \caption{Fast path execution for independent objects. Two clients (C1, C2) write to different objects (ObjA, ObjB) concurrently. Coordinators R1 and R7 broadcast proposals to all replicas during the propose phase. Dotted lines indicate weighted responses from high-weight replicas. Both objects commit in parallel.}
\label{fig:fastpath-msg}
\Description{Sequence diagram showing parallel fast-path execution with two concurrent writes to independent objects, illustrating propose, reply, and commit phases.}

\end{figure}
The figure illustrates the fast path execution where two clients concurrently write to different independent objects (Client1 writes ObjA, Client2 writes ObjB). The nodes (R1 and R7) receive these write requests, they would be the temporary coordinator for this request and broadcast them to all replicas (R1-R7) during the propose phase. Since there's no conflict between the operations, replicas can independently process and acknowledge the requests during the reply phase. The dotted lines represent acknowledgments flowing back to the coordinators, with different weights indicating replica responses. Once sufficient acknowledgments are received, the coordinators can commit their respective objects (ObjA and ObjB) without additional coordination, allowing both transactions to complete in parallel.

\begin{algorithm}[t]
\caption{Fast Path Protocol}
\label{algo:fastpath}
\begin{algorithmic}[1]
\Procedure{FastPath}{$op, O$}
    \If{$O$ has conflicting operation in progress}
        \State \Return \Call{SlowPath}{$op, O$}
    \EndIf
    \State $weight \gets w_{\text{self}}^O$
    \For{each replica $r$}
        \State send \texttt{FAST\_PROPOSE}$(op, O)$ to $r$
    \EndFor
    \While{$weight < T^O$ \textbf{and} not timeout}
        \State $msg \gets$ receive()
        \If{$msg =$ \texttt{FAST\_ACCEPT} from $r$}
            \State $weight \gets weight + w_r^O$
            \If{$weight \geq T^O$}
                \State broadcast \texttt{FAST\_COMMIT}$(op)$
                \State \Return \textsc{Success}
            \EndIf
        \ElsIf{$msg =$ \texttt{CONFLICT}}
            \State \Return \Call{SlowPath}{$op, O$}
        \EndIf
    \EndWhile
    \State \Return \Call{SlowPath}{$op, O$}
\EndProcedure
\end{algorithmic}
\end{algorithm}

\subsection{Slow Path: Node-Weighted Consensus}
\begin{algorithm}[t]
\caption{Slow Path Protocol}
\label{algo:slowpath}
\begin{algorithmic}[1]
\Procedure{SlowPath}{$op, O$}
    \If{not leader}
        \State forward to leader
    \EndIf
    \State lock(mutex)
    \State $priority \gets$ getPriorities()
    \State $pSum \gets p_{\text{self}}$
    \For{each follower $f$}
        \State send \texttt{SLOW\_PROPOSE}$(op, priority)$ to $f$
    \EndFor
    \While{$pSum < T^N$ \textbf{and} not timeout}
        \State $msg \gets$ receive()
        \If{$msg =$ \texttt{SLOW\_ACCEPT} from $f$}
            \State $pSum \gets pSum + p_f$
            \If{$pSum \geq T^N$}
                \State broadcast \texttt{SLOW\_COMMIT}$(op)$
                \State updatePriorities(responders)
                \State unlock(mutex)
                \State \Return \textsc{Success}
            \EndIf
        \EndIf
    \EndWhile
    \State unlock(mutex)
    \State \Return \textsc{Failure}
\EndProcedure
\end{algorithmic}
\end{algorithm}

The slow path handles \textbf{conflicting or shared objects} via leader-coordinated node-weighted consensus. Algorithm~\ref{algo:slowpath} presents the complete slow-path protocol:

\begin{enumerate}
\item \textbf{Leader Forwarding:} Coordinators forward operations to the current leader (lines 2-4).

\item \textbf{Priority Assignment:} The leader maintains a priority clock and assigns priority weights to replicas based on recent response latencies (line 6).

\item \textbf{Slow-Path Proposal:} The leader broadcasts \texttt{SLOW\_PROPOSE} including operation and priority information (lines 8-10).

\item \textbf{Priority-Weighted Voting:} Followers respond with (line 13)  \texttt{SLOW\_ACCEPT} ; the leader accumulates priority weights (line 14).

\item \textbf{Quorum Formation and Commit:} Once accumulated priority exceeds node quorum threshold $T^N$ (line 15), the leader commits and broadcasts \texttt{SLOW\_COMMIT} (line 16).
\end{enumerate}

\begin{figure}
    \centering
    \includegraphics[width=0.99\linewidth]{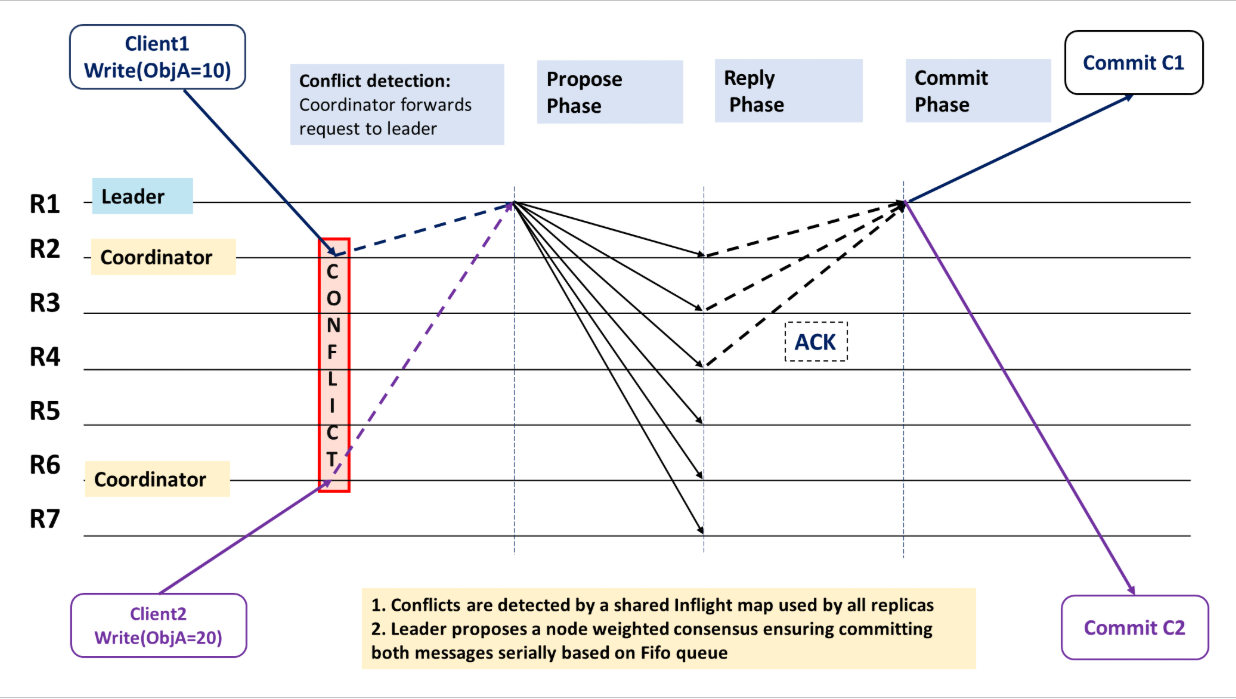}
    \caption{Slow path conflict resolution for shared objects. Two clients write conflicting values to the same object (ObjA). Conflict detection triggers forwarding to leader R1, which serializes the operations using node-weighted consensus. The FIFO queue ensures sequential commits (C1 then C2) maintaining linearizability.}
\label{fig:slowpath-msg}
\Description{Sequence diagram showing conflict detection and leader-coordinated slow-path resolution with serialized commits for conflicting writes.}

\end{figure}
The figure shows the conflict resolution mechanism when concurrent writes create a conflict (Client1 writes ObjA=10, Client2 writes ObjA=20, both targeting the same object). When the conflict is detected through a shared in-flight map maintained by all replicas, the coordinator (R2 and R6) forwards the conflicting requests to a designated leader (R1). The red "CONFLICT" box indicates where the conflict detection occurs. The leader then proposes a serialization order using node-weighted consensus, ensuring both conflicting writes are committed serially based on a FIFO queue. This guarantees consistency by forcing one operation to complete before the other, with all replicas acknowledging the leader's proposed ordering before final commitment, resulting in sequential commits (c1 then c2) rather than parallel execution.
\subsection{Correctness Properties}
WOC guarantees linearizability through weighted quorum intersection, formalized by the following invariants and theorems.

\textbf{Invariants.} WOC inherits Cabinet’s safety and liveness invariants at object granularity.

\textit{Invariant I1.} For any object $O$, the sum of weights of the top $t+1$ replicas exceeds the consensus threshold:
$$
\sum_{i=1}^{t+1} w_i^O > T^O = \frac{1}{2}\sum_{j=1}^n w_j^O
$$

\textit{Invariant I2 (Safety).} Any set of $t$ replicas cannot form a quorum:
$$
\forall S, |S|=t:\quad \sum_{i \in S} w_i^O < T^O
$$

Together, I1 and I2 ensure fault tolerance: progress is possible despite up to $t$ failures, while every decision requires participation from at least $t+1$ replicas.

\textbf{Theorem 1 (Fast-Path Safety).} \textit{If two operations $op_1$ and $op_2$ on object $O$ commit via the fast path, their quorums $Q_1^O$ and $Q_2^O$ intersect.}

\begin{proof}
By contradiction. Assume $Q_1^O \cap Q_2^O = \emptyset$. Since both operations commit,
$$
\sum_{i \in Q_1^O} w_i^O \ge T^O \quad \text{and} \quad \sum_{j \in Q_2^O} w_j^O \ge T^O .
$$
Summing gives
$$
\sum_{i \in Q_1^O} w_i^O + \sum_{j \in Q_2^O} w_j^O \ge 2T^O = \sum_{k=1}^n w_k^O .
$$
Because the quorums are disjoint, this exceeds the total system weight, a contradiction. Hence $Q_1^O \cap Q_2^O \neq \emptyset$.
\end{proof}

\textbf{Theorem 2 (Cross-Path Consistency).} \textit{Operations on the same object cannot commit concurrently via different paths.}

\textit{Proof sketch.} The object manager tracks in-flight operations per object (Algorithm~1). Fast-path requests check this state and defer to the slow path if a conflicting operation exists, while slow-path operations acquire a mutex, preventing concurrent fast-path commits.

\subsubsection{Liveness Guarantee}

\textbf{Liveness Property.} \textit{If the top $t+1$ weighted replicas remain responsive, every operation eventually commits.}

Fast-path operations either commit upon receiving sufficient responses or redirect to the slow path on conflict or timeout. The slow path inherits Cabinet’s liveness guarantees, ensuring progress as long as the leader and $t$ other replicas remain available. Thus, operations never block indefinitely.

%% file: sections/Evaluation.tex
\section{Evaluation}
\label{sec:evaluation}

We implemented \WOC in Go and evaluated its performance against Cabinet on cloud infrastructure, assessing its dual-path approach under mixed workloads reflecting independent, common, and conflicting object accesses.
\begin{figure*}[!t]
\centering
\begin{minipage}{0.31\textwidth}
    \centering
    \includegraphics[width=\linewidth]{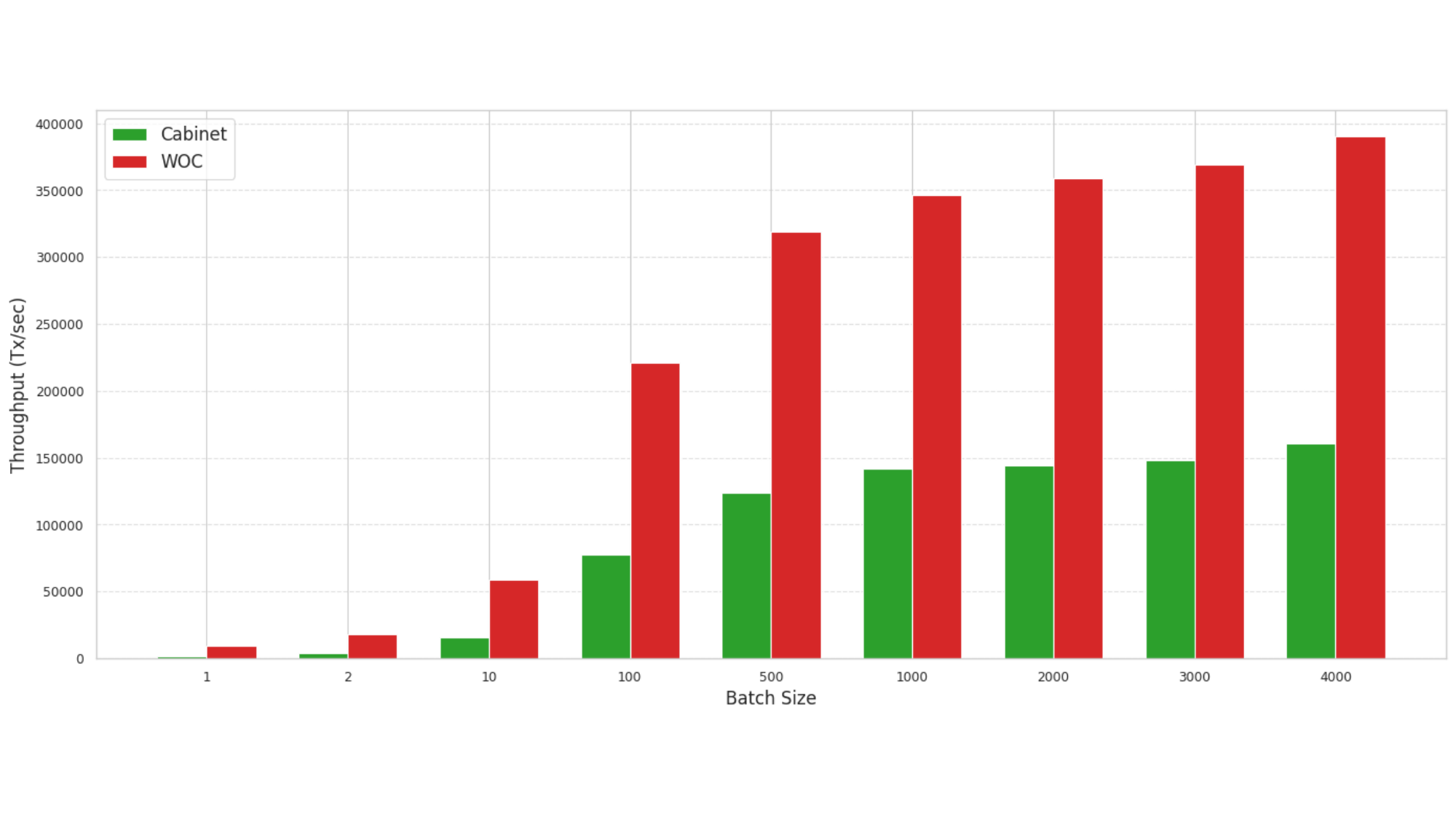}
    \subcaption{Throughput vs Batch Size}
    \label{fig:batchscale_tpt}
\end{minipage}\hfill
\begin{minipage}{0.31\textwidth}
    \centering
    \includegraphics[width=\linewidth]{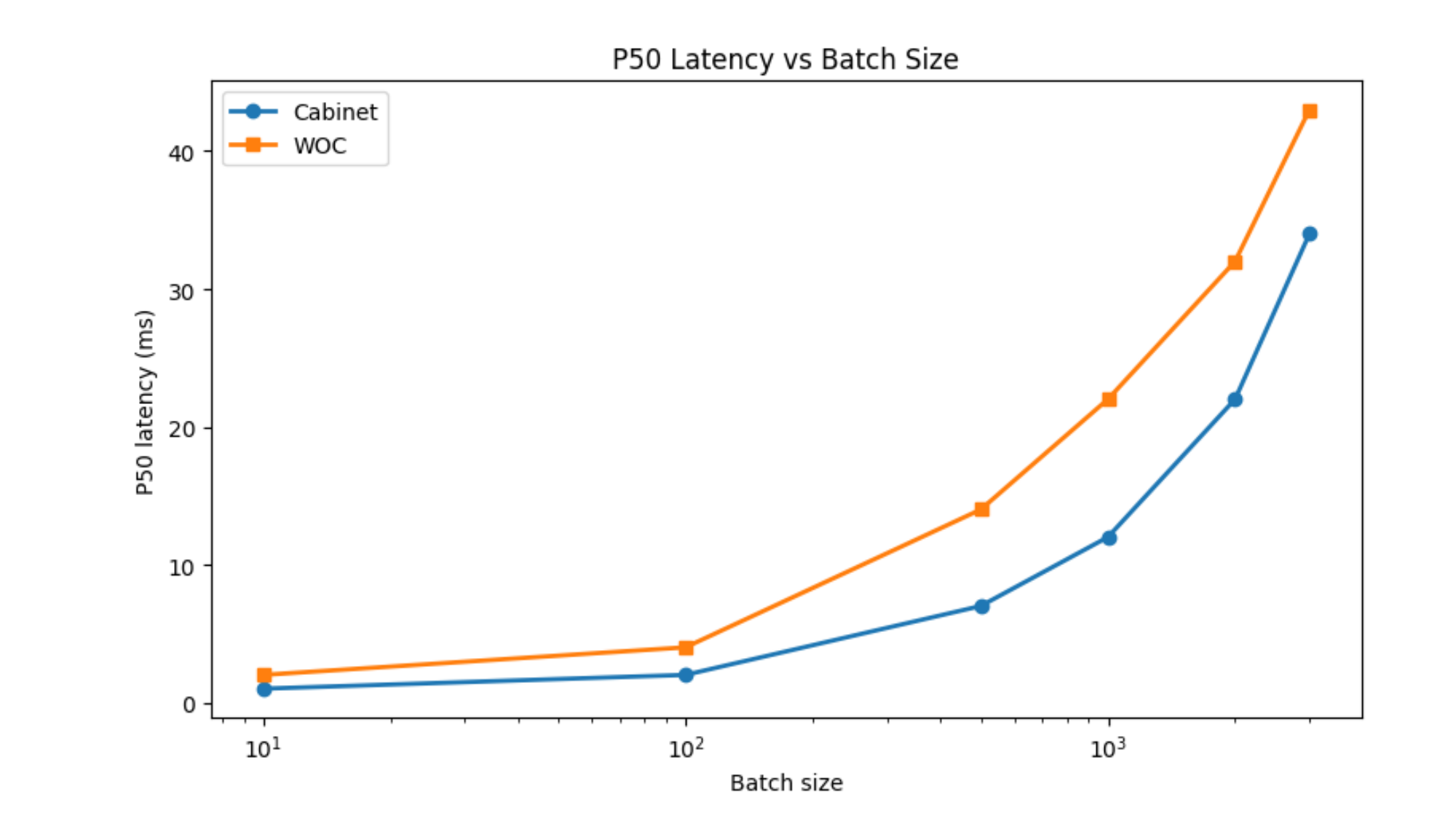}
    \subcaption{P50 Latency vs Batch Size}
    \label{fig:batchscale_p50}
\end{minipage}\hfill
\begin{minipage}{0.31\textwidth}
    \centering
    \includegraphics[width=\linewidth]{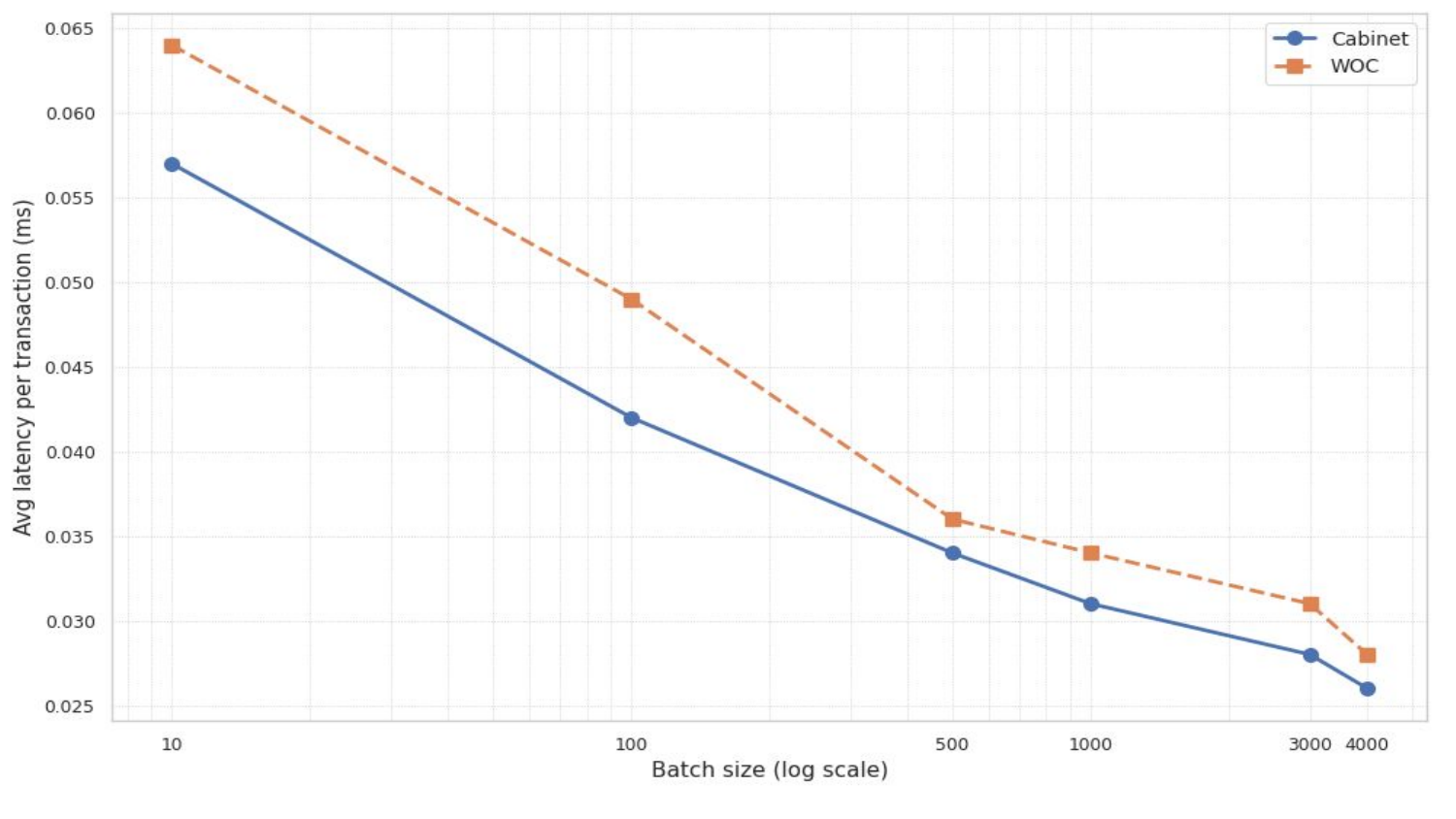}
    \subcaption{Average Latency vs Batch Size}
    \label{fig:batchscale_avg}
\end{minipage}
\caption{Performance under varying batch sizes (1 to 4000 operations)
(a) \WOC achieves up to 3× higher throughput than Cabinet for small to medium batches, exceeding 300k Tx\/s at batch size 1000 due to parallel object-based commits, while Cabinet plateaus around 160k Tx\/s due to leader serialization.
(b) P50 latency versus batch size: WOC exhibits higher median latency as a trade-off for increased throughput; both systems show saturation beyond batch size 2000.
(c) Average latency versus batch size: average latency trends mirror the median behavior, with \WOC incurring higher latency under larger batches as contention and batching overhead increase.}

\label{fig:batchscale}
\Description{Three graphs showing throughput and latency trends as batch size increases, comparing WOC's parallel execution advantage against Cabinet's leader-limited performance.}
\end{figure*}

\subsection{Experimental Setup}
Our experimental deployment uses Compute Canada cloud infrastructure with a baseline configuration of 5 replica servers and 2 client VMs. The default workload distribution comprises 90\% independent objects (fast-path eligible), 5\% common objects (occasional conflicts), and 5\% conflicting objects (hot objects requiring slow-path coordination), with each operation carrying a 512-byte payload.
Clients employ open-loop request generation with a maximum of 5 in-flight requests to prevent overload. \WOC distributes requests round-robin across available replicas, enabling parallel request ingestion, while Cabinet routes all requests to a single leader replica. This setup simulates realistic object access patterns in multi-tenant environments.
We benchmark \WOC and Cabinet along four dimensions:
\begin{itemize}
    \item \textbf{Batch size scaling:} Varying batch sizes from 10 to 6000 operations with 5 servers and 2 clients
    \item \textbf{Conflict rate sensitivity:} Testing conflict rates from 0\% to 100\% with 5 servers and 2 clients
    \item \textbf{Client concurrency:} Scaling from 2 to 9 concurrent clients with 5 servers
    \item \textbf{Server scalability:} Scaling from 3 to 9 replica servers with 2 clients
\end{itemize}
All experiments use 512-byte messages and maintain identical hardware configurations for fair comparison. Server configurations tolerate  $f = 2$ failures,where n is the number of replicas.

\subsection{Batch Size Scaling}
We evaluate the impact of increasing batch size on throughput and latency for \WOC and Cabinet, varying batch sizes from 1 to 4000. Figures~\ref{fig:batchscale_tpt}, \ref{fig:batchscale_p50}, and \ref{fig:batchscale_avg} show the throughput, P50 latency, and average latency characteristics, respectively.
As batch size grows, \WOC consistently outperforms Cabinet due to its distributed request handling and fast-path execution for independent objects. For small batches (10-100 operations), \WOC achieves approximately 5x times
 higher throughput than Cabinet, reaching 9.1k--17.6k Tx/sec compared to Cabinet's 1.8k--3.5k Tx/sec. Cabinet's single-leader architecture funnels all requests through a sequential bottleneck, fundamentally limiting its scalability.

As batch sizes reach 500--1000 operations, \WOC enters a high-throughput regime, scaling from 58.5k to 220.8k Tx/sec, while Cabinet reaches 15.3k to 77.8k Tx/sec. This performance gap reflects \WOC's ability to leverage parallel object-based commits and reduced coordination overhead per transaction. At larger batch sizes (2000--4000), \WOC sustains 319k--390k Tx/sec, maintaining a 
3× throughput advantage over Cabinet, which plateaus at 123k--161k Tx/sec due to leader bottlenecks.

In terms of latency, \WOC exhibits higher P50 latencies than Cabinet due to coordination across different paths ranging from 2ms at batch size 10 to 43ms at batch size 3000, compared to Cabinet's 1ms to 34ms. However, average latencies remain low for both systems (0.028--0.064ms for \WOC vs. 0.026--0.057ms for Cabinet), and this latency trade-off enables substantially higher throughput. The results demonstrate that \WOC's distributed architecture effectively exploits batch parallelism while Cabinet's centralized design limits scalability despite batching optimizations.
\begin{figure*}[!t]
\centering
\begin{minipage}{0.48\textwidth}
    \includegraphics[width=\linewidth]{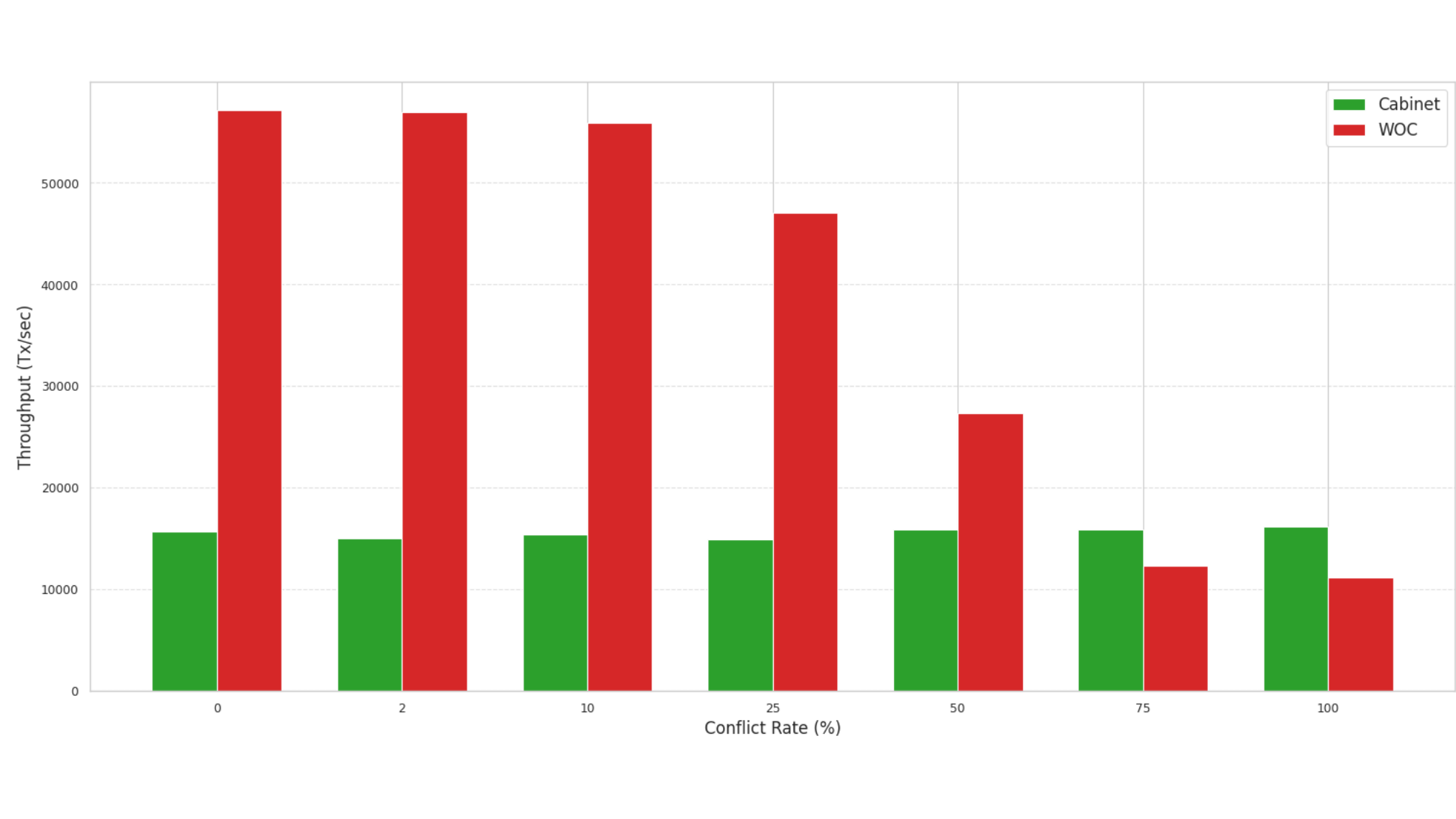}
    \subcaption{Throughput vs Conflict Rate}
    \label{fig:conftpt}
\end{minipage}\hfill
\begin{minipage}{0.48\textwidth}
    \includegraphics[width=\linewidth]{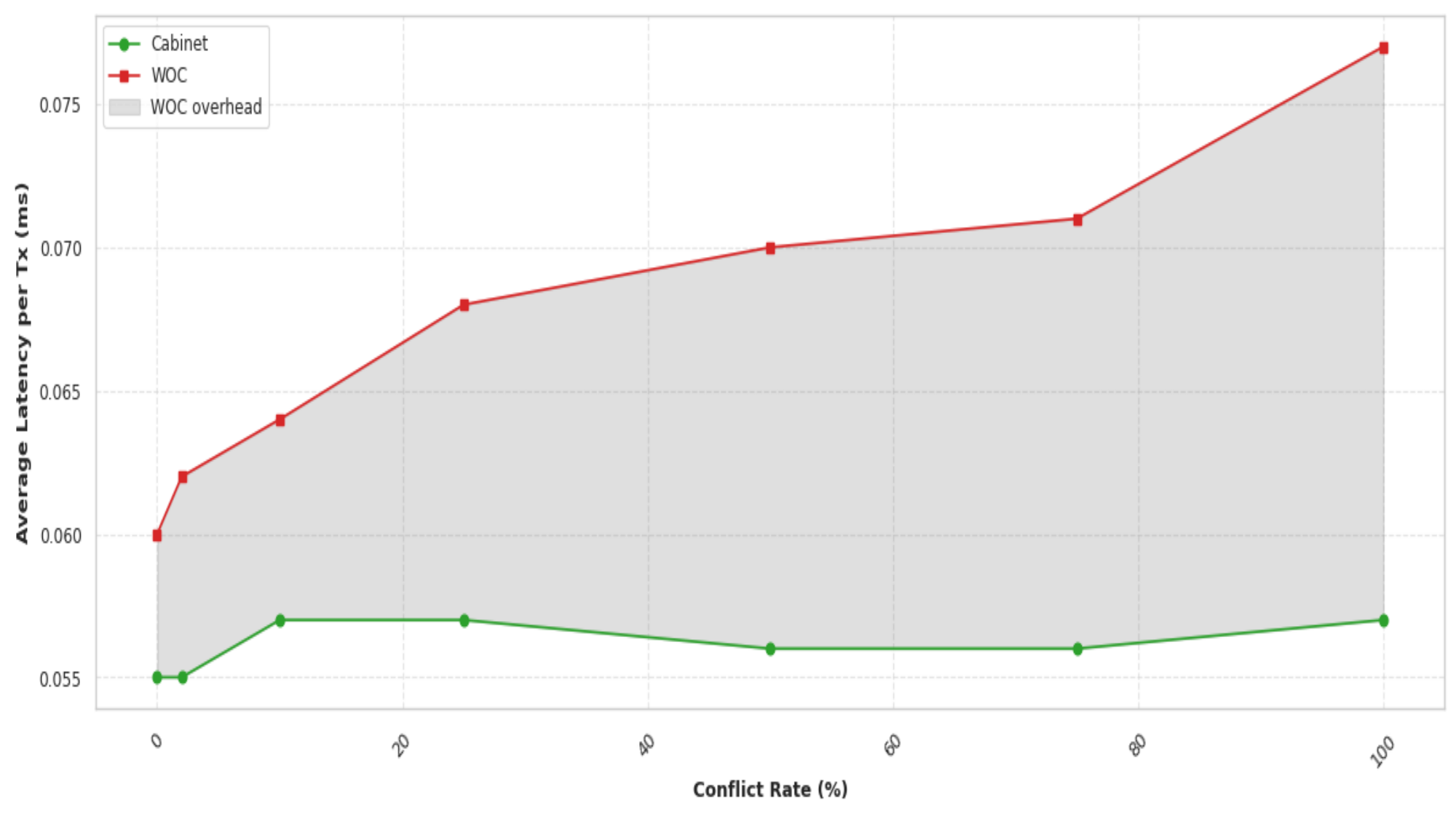}
    \subcaption{Latency vs Conflict Rate}
    \label{fig:conflat}
\end{minipage}
\caption{Performance under varying conflict rates (0\% to 100\%, batch size 10). (a) WOC maintains 55-57k Tx/sec for 0-10\% conflicts with >95\% fast-path commits, but degrades to 12k Tx/sec at 100\% conflict. Cabinet provides stable 14-16k Tx/sec across all conflict levels. Crossover occurs at 60-70\% conflict rate. (b) WOC latency increases with conflict rate (0.2ms to 0.38ms) while Cabinet remains constant (0.03-0.06ms).}
\label{fig:conflict}
\Description{Two graphs demonstrating WOC's superior performance in low-conflict scenarios versus Cabinet's consistent performance across all contention levels.}
\end{figure*}

\subsection{Conflict Rate Scaling}
We evaluate the impact of varying conflict percentages while keeping batch size fixed at 10, testing rates of 0\%, 2\%, 10\%, 25\%, 50\%, 75\%, and 100\%. Figure~\ref{fig:conflict} illustrates the performance trends across throughput and latency metrics.
\WOC demonstrates high efficiency under low and medium-conflict workloads. For 0--10\% conflicts, operations predominantly commit via the fast path, achieving stable throughput of approximately 55.9k--57.1k Tx/sec roughly 3.8× higher than Cabinet's 14.9k--15.7k Tx/sec. Average latencies remain low for both systems, with \WOC at 0.060--0.064ms and Cabinet at 0.055--0.057ms. The fast path's leaderless coordination allows concurrent execution of independent operations across multiple replicas, effectively multiplying the system's processing capacity.

As conflict rates rise to 25\%, fast-path commits decrease, reducing \WOC's throughput to 47.1k Tx/sec while maintaining a 3.2× advantage over Cabinet (14.9k Tx/sec). At 50\% conflict, \WOC's throughput drops to 27.3k Tx/sec but still outperforms Cabinet's 15.9k Tx/sec by 1.7×. However, at 75--100\% conflict, \WOC's throughput falls to 11.2k--12.3k Tx/sec, now underperforming Cabinet's stable 15.9k--16.2k Tx/sec. Latency increases modestly for \WOC (0.071--0.077ms) while Cabinet remains consistent (0.056--0.057ms).

In contrast, Cabinet exhibits minimal sensitivity to conflict levels because all operations are serialized through a single global leader. Its throughput remains relatively stable between 14.9k--16.2k Tx/sec across all conflict percentages with consistently low latencies. The crossover point occurs around 75\% conflict rate, beyond which Cabinet's consistent performance surpasses \WOC's degraded throughput as the benefits of distributed coordination diminish under high contention.

\begin{figure*}[!t]
\centering
\begin{minipage}{0.48\textwidth}
    \includegraphics[width=\linewidth]{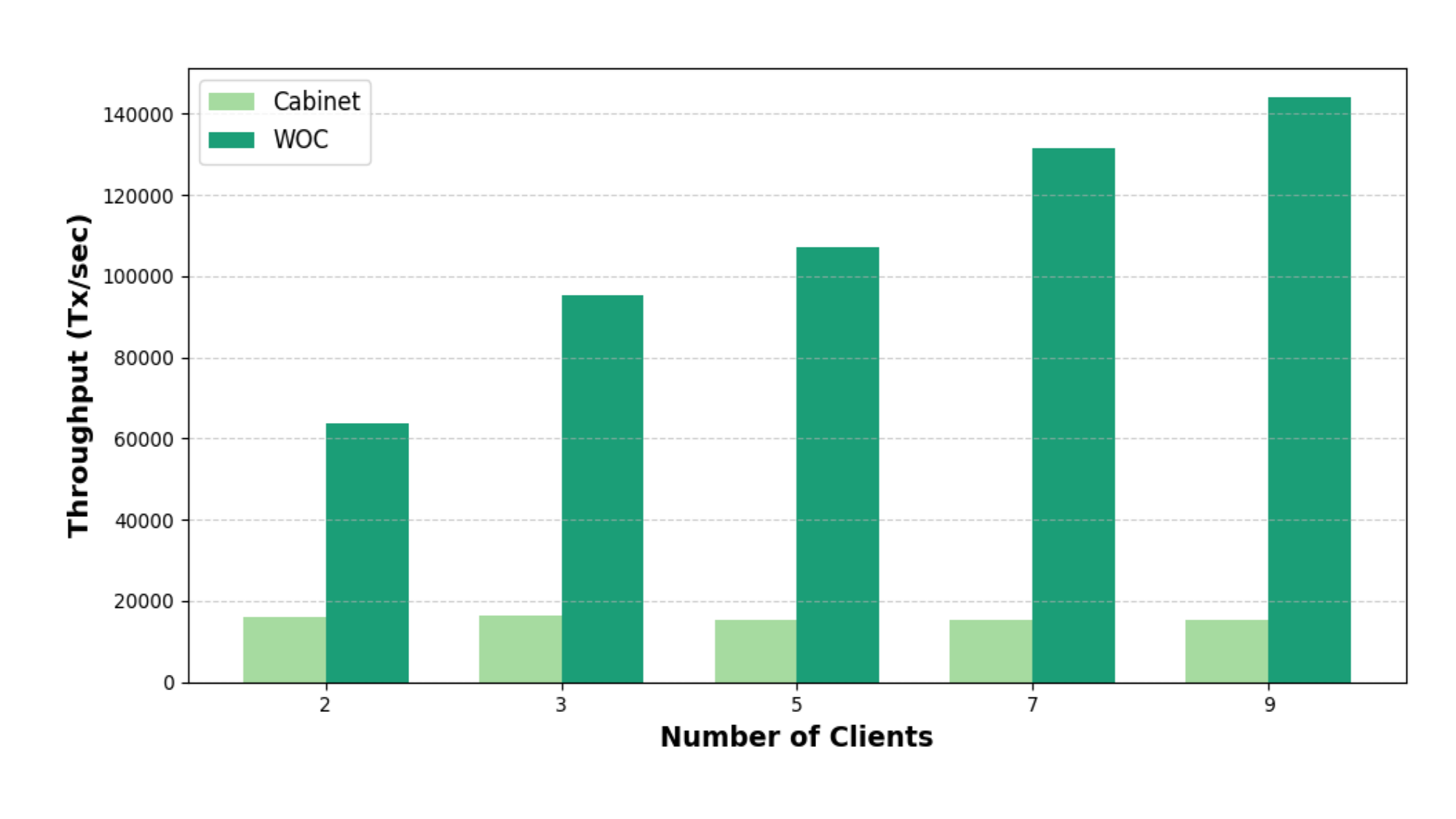}
    \subcaption{Throughput vs Number of Clients}
    \label{fig:clienttpt}
\end{minipage}\hfill
\begin{minipage}{0.48\textwidth}
    \includegraphics[width=\linewidth]{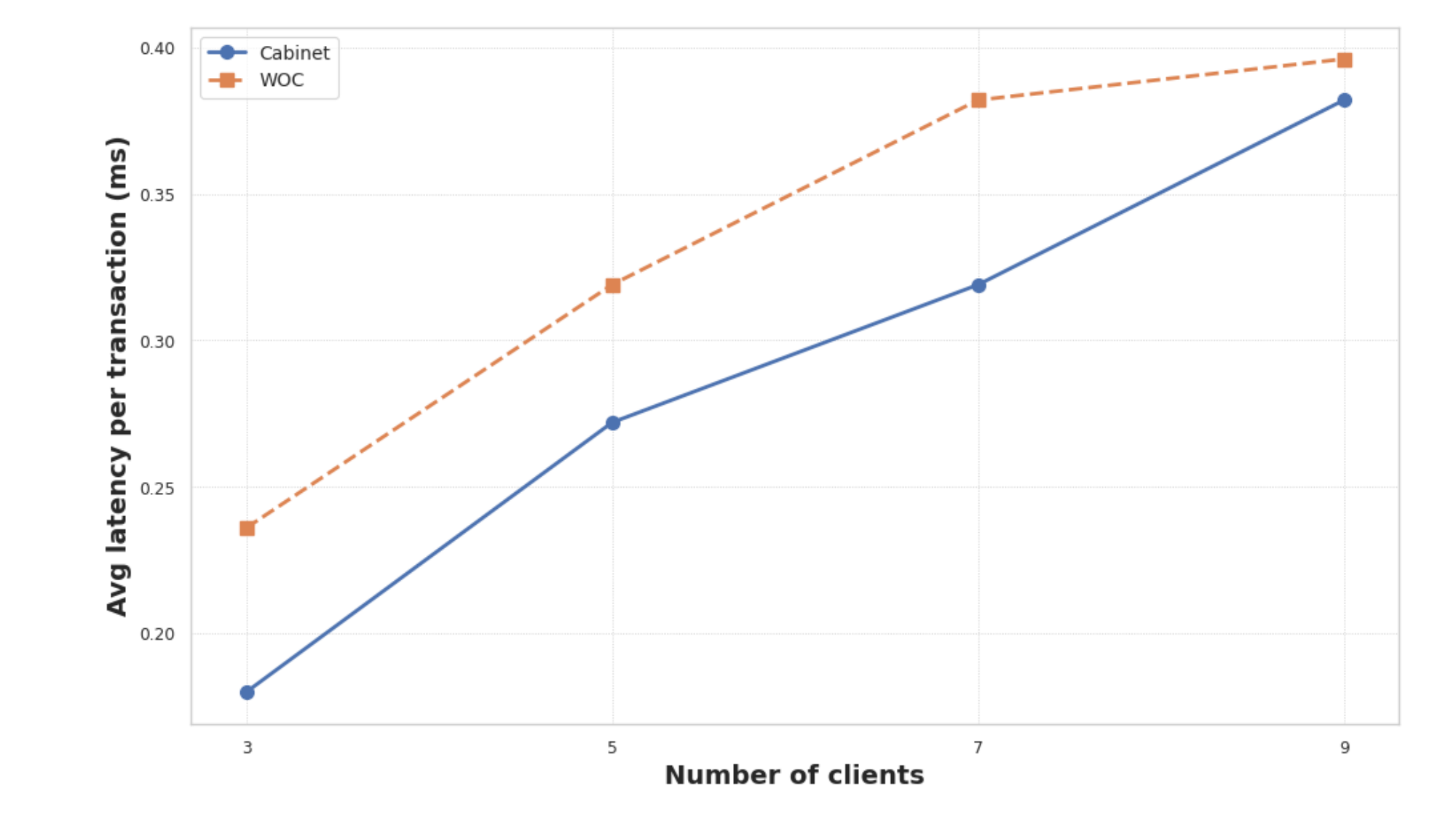}
    \subcaption{Latency vs Number of Clients}
    \label{fig:clientlat}
\end{minipage}
\caption{Performance under increasing client concurrency (2 to 9 clients). (a) WOC scales nearly linearly from 2 to 7 clients, reaching 144k Tx/sec at 9 clients through distributed request ingestion. Cabinet plateaus at 15-16k Tx/sec due to single-leader bottleneck. (b) WOC latency increases slightly at 9 clients (0.225ms) due to queuing delays, while Cabinet maintains low latencies (0.03-0.06ms).}
\label{fig:clientscale}
\Description{Two graphs showing WOC's effective load distribution across replicas versus Cabinet's leader saturation as client count increases.}
\end{figure*}
\subsection{Client Scaling}
We evaluate performance as concurrent clients increase, testing 2, 3, 5, 7, and 9 clients. Figure~\ref{fig:clientscale} shows the scaling characteristics across throughput and latency dimensions.

\WOC exhibits strong throughput scaling from 2 to 9 clients, leveraging decentralized request ingestion and efficient batching pipelines. Throughput grows from 63.6k Tx/sec at 2 clients to 131.5k Tx/sec at 7 clients, reaching a peak of 144.1k Tx/sec at 9 clients—a $2.3\times$ improvement. The distributed nature of \WOC's fast path allows multiple coordinators to process independent operations simultaneously, with each replica acting as a coordinator for fast-path operations and distributing coordination load across the cluster.

The linear scaling up to 7 clients (from 63.6k to 131.5k Tx/sec) demonstrates that \WOC's dual-path architecture successfully distributes load across replicas. Beyond 7 clients, throughput growth moderates as the system approaches resource saturation on individual replicas, particularly CPU utilization for message processing and quorum computation. Average latency increases from 0.236ms at 3 clients to 0.396ms at 9 clients, reflecting queuing delays.

In contrast, Cabinet scales minimally due to its single-leader implementation. Throughput remains essentially flat between 15.4k--16.3k Tx/sec across all client counts, representing only $9$--$11\%$ of \WOC's peak performance. Average latency increases from 0.180ms at 3 clients to 0.382ms at 9 clients, indicating that the leader becomes saturated and cannot process incoming requests fast enough despite increased client concurrency. Cabinet's flat scaling curve confirms that its centralized architecture cannot benefit from additional client parallelism once the leader is saturated, fundamentally limiting scalability.
\begin{figure*}[!t]
\centering
\begin{minipage}{0.48\textwidth}
    \includegraphics[width=\linewidth]{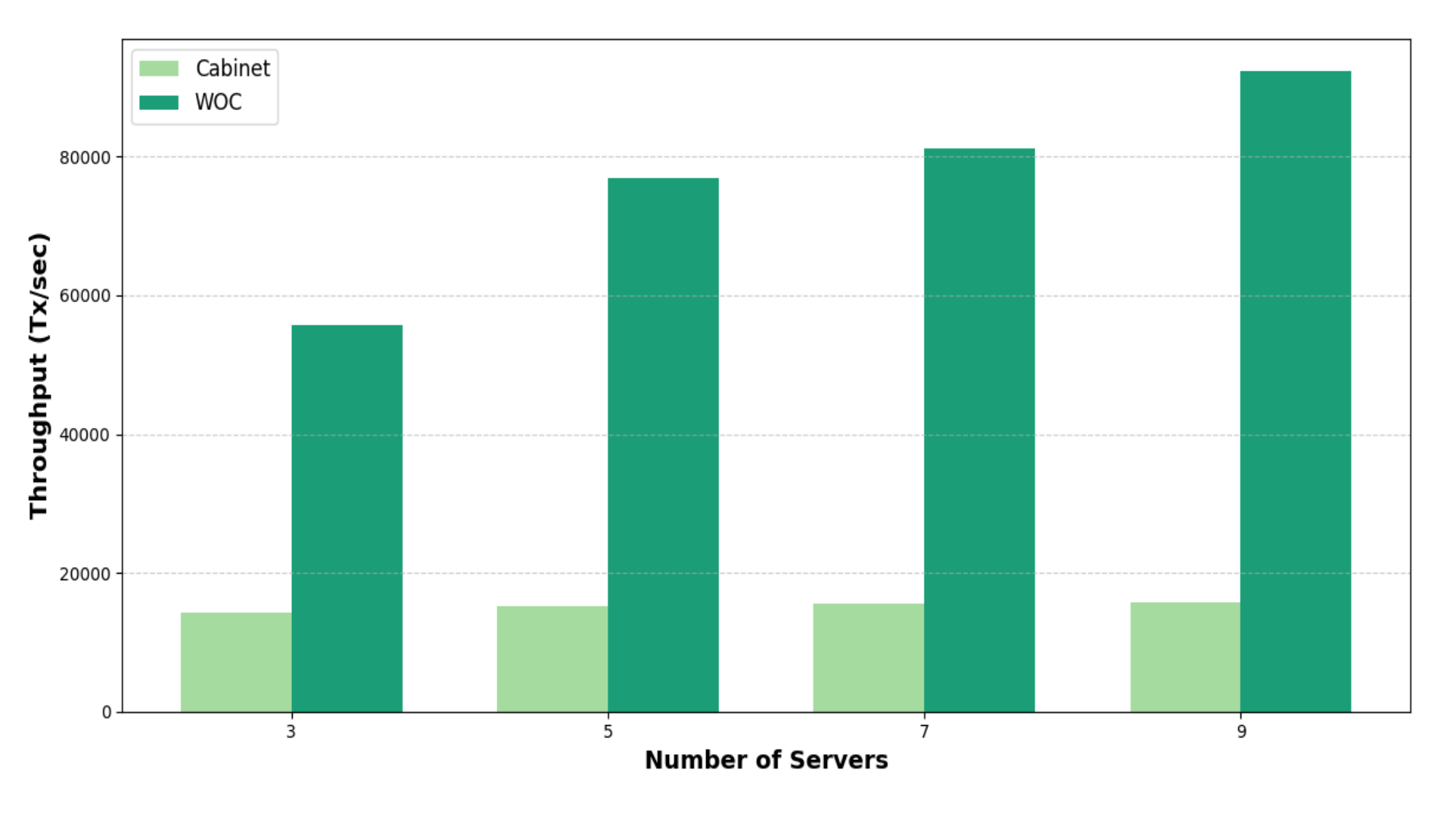}
    \subcaption{Throughput vs Number of Servers}
    \label{fig:servertpt}
\end{minipage}\hfill
\begin{minipage}{0.48\textwidth}
    \includegraphics[width=\linewidth]{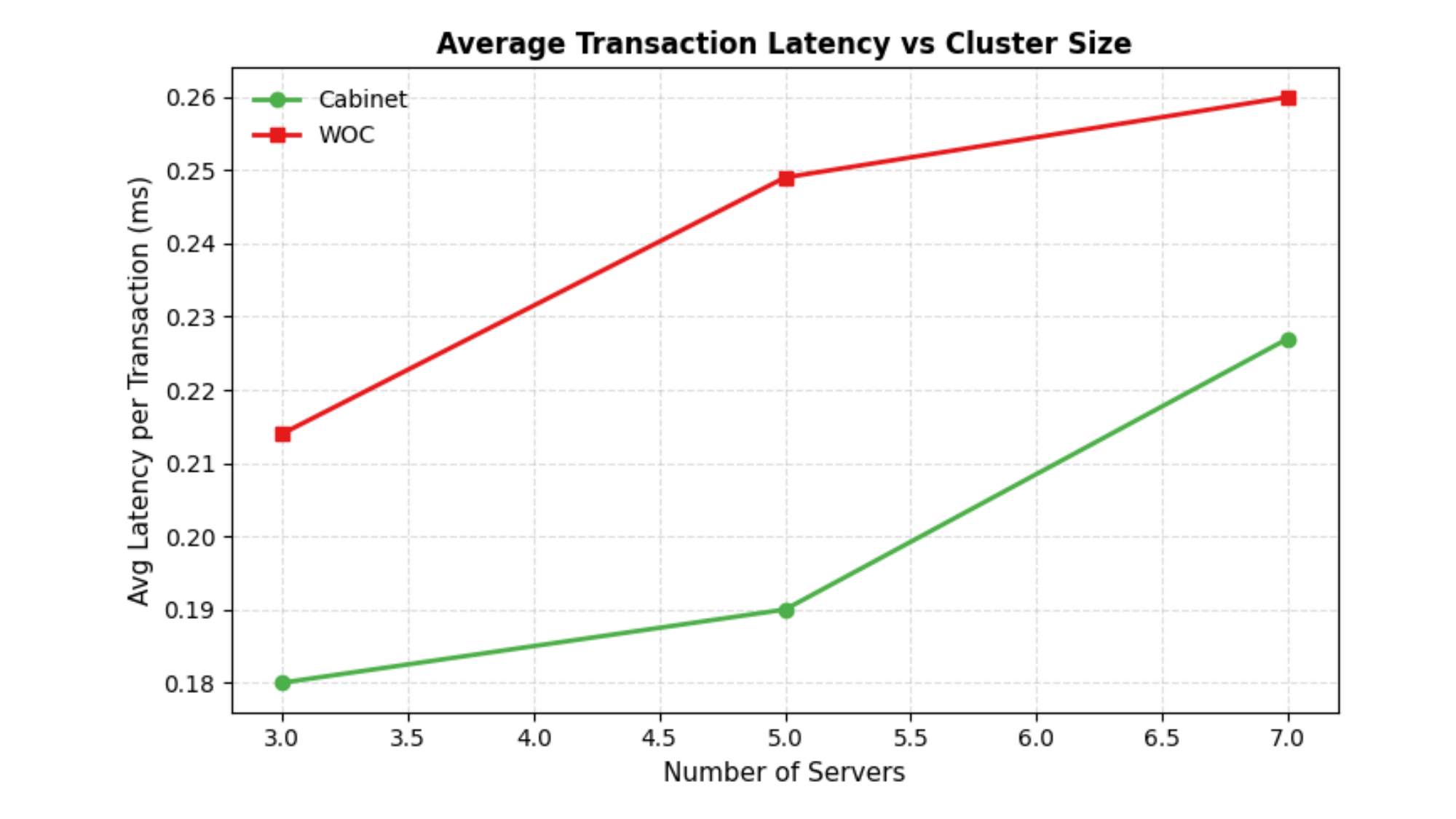}
    \subcaption{Latency vs Number of Servers}
    \label{fig:serverlat}
\end{minipage}
\caption{Performance under increasing server count (3 to 9 replicas). (a) WOC scales from 55k to 81k Tx/sec (3 to 7 servers) through parallel request ingestion, then tapers beyond 7 servers due to coordination overhead. Cabinet improves modestly from 14k to 15.7k Tx/sec. (b) WOC latency increases slightly with fan-out costs (0.135ms to 0.237ms), while Cabinet latency decreases due to faster quorum formation.}
\label{fig:serverscale}
\Description{Two graphs showing WOC's strong positive scaling compared to Cabinet's limited improvement as cluster size increases.}
\end{figure*}
\subsection{Server Scaling}
We evaluate performance while scaling replica servers from 3 to 9, testing configurations of 3, 5, 7, and 9 servers. Figure~\ref{fig:serverscale} presents the scaling behavior across throughput and latency dimensions.
\WOC demonstrates strong positive scaling throughout the entire range, with throughput increasing from 55.8k Tx/sec at 3 servers to 92.4k Tx/sec at 9 servers, 1.66× improvement. The system exhibits particularly strong gains from 3 to 5 servers (55.8k to 77.0k Tx/sec), driven by parallel request and reduced contention on each replica's message-processing pipeline. The fast path benefits from having more replicas available to act as coordinators, while distributed quorum formation allows operations to complete without waiting for all replicas.

Throughput continues to scale from 7 to 9 servers (81.3k to 92.4k Tx/sec), demonstrating that \WOC can effectively leverage additional replicas even in larger cluster configurations. Average latency increases from 0.214ms at 3 servers to 0.260ms at 7 servers, reflecting the increased coordination overhead and higher fan-out costs during cross-replica validation in larger clusters.
In contrast, Cabinet exhibits minimal scaling benefits as the cluster grows. Throughput rises only from 14.3k Tx/sec at 3 servers to 15.7k Tx/sec at 9 servers, a mere 1.1× improvement. Average latency also increases from 0.180ms to 0.227ms as cluster size grows, indicating higher coordination costs. Cabinet's throughput improvement is primarily due to reduced contention on the leader's message processing pipeline and faster quorum formation, rather than increased parallelism.

Overall, \WOC maintains a 3.5× throughput advantage over Cabinet across all cluster sizes, demonstrating that its distributed ingestion and fast-path commits scale more effectively than Cabinet's centralized leader architecture.

\subsection{Summary of Results}
Across all evaluations — batch size, conflict rate, client concurrency, and server scaling \WOC consistently outperforms Cabinet in throughput for workloads dominated by independent or low-to-moderate conflict objects. The fast-path execution enables \WOC to efficiently exploit batching, parallel requests, and distributed coordination across multiple replicas, yielding substantial performance gains.
For low-conflict workloads, \WOC achieves several times higher throughput than Cabinet while maintaining sub-millisecond latencies. This advantage persists across batch size scaling, where \WOC demonstrates superior parallelization, and client scaling, where \WOC's distributed architecture effectively leverages increased concurrency while Cabinet's single-leader bottleneck limits scalability. Server scaling further confirms these advantages, with \WOC showing strong throughput improvements as cluster size increases, while Cabinet exhibits minimal gains due to its centralized design.

The performance crossover occurs only at very high conflict rates, where Cabinet's stable performance surpasses \WOC's degraded throughput. This demonstrates that Cabinet's single-leader, weighted quorum design provides consistent performance under extreme contention, where conflict detection overhead negates \WOC's parallelization benefits. The latency trade-off is modest, with \WOC exhibiting slightly higher average latencies: a reasonable cost for substantial throughput advantages.
These results validate \WOC's dual-path design philosophy: leveraging the fast path for efficiency under typical workloads while maintaining correctness through the slow path under high contention. The evaluation confirms that combining object-level and node-level weighted consensus provides significant performance advantages for modern distributed systems with heterogeneous, predominantly low to medium conflict workloads.

%% file: sections/Conclusion.tex
\section{Conclusions}
This paper presents \WOC, a consensus protocol that integrates object-weighted and node-weighted consensus to efficiently handle heterogeneous workloads in distributed systems. By providing a leaderless fast path for independent objects and a priority-based slow path for conflicting objects, \WOC achieves high throughput, outperforming Cabinet by $3$--$5\times$ under low-conflict workloads while maintaining robust performance under contention.

Our results highlight that combining object- and node-level weighted consensus can effectively balance efficiency and safety, offering a promising direction for scalable, adaptive consensus in geographically distributed systems. Future work includes exploring adaptive threshold tuning, integrating Byzantine fault tolerance, and evaluating \WOC in production geo-distributed environments.